# Hong-Ou-Mandel interference between two independent all fiber photon sources

Zhi-Yuan Zhou,[1,2][†] Yin-Hai Li,[3][†] Zhao-Huai Xu,[3] Shuang Wang,[1,2] Li-Xin Xu,[3*] Bao-Sen Shi,[1,2][‡]

Guang-Can Guo[1,2]

[1]*Key Laboratory of Quantum Information, University of Science and Technology of China, Hefei, Anhui 230026, China*

[2]*Synergetic Innovation Center of Quantum Information & Quantum Physics, University of Science and Technology of China,*

*Hefei, Anhui 230026, China*

[3]*Department of Optics and Optical Engineering, University of Science and Technology of China, Hefei, Anhui 230026, China*

*[*]xulixin@ustc.edu.cn*

*[‡]drshi@ustc.edu.cn*

Guided-wave platforms such as fiber and silicon-on-insulator waveguide show great advances over traditional free space implementations in quantum information technology for significant advantages of low transmission loss, low cost, integrability and compatible with mature fiber communication systems. Interference between independent photon sources is the key to realize complex quantum systems for more sophisticated applications such as multi-photon entanglement generation and quantum teleportation. In this work, we report Hong-Ou-Mandel interference between two independent all fiber photon pair sources over two 100GHz dense wave division multiplexing channels, the visibility reaches 53.2±8.4% (82.9±5.3%) without (with) back ground counts subtracted. In addition, we give a general theoretical description of the purity of the photon pair generation in dispersion shifted fiber and obtain the optimized condition for high purity photon pair generation. We also obtain a maximum coincidence to back ground ratio of 131 by cooling the fiber in liquid nitrogen. Our study shows great promising of integrated optical elements for future scalable quantum information promising.

Interference plays a vital important role in both classical and quantum optical technology, which has many fundamental applications such as high precision optical metrology [1], generation of multi-photon entangled states for quantum computation [2], and quantum teleportation [3], device independent quantum key distributions for quantum communications [4, 5]. High visibility non-classical interference between independent single photon is the key technique for generating complex multi-photon GHZ state[6, 7], cluster state [8] and high photon number NOON states[9], these states are elemental quantum resources for harnessing quantum technology. Therefore non-classical interference between independent single photon sources has been widely studied in literature. Original studies are in the visible wavelength regime, which based on photon sources generated from spontaneous parametric down conversion (SPDC) in second order nonlinear crystals [10-13] or spontaneous four-wave-mixing (SFWM) in photonic crystal fibers[14-16]. Only recently, investigation in telecom band is realized with photon sources from SPDC in nonlinear crystal [17, 18]. For all these experiments, the key point to increase the interference visibility is to increase the purity of the single photon source by either engineering the dispersion and phase match condition or using narrow bandwidth filters.

---

[†]**These two authors contributed equally to this article.**

Nowadays, single photon sources based on guided-wave platforms such as dispersion-shifted fibers (DSF) and silicon-on-insulator (SOI) waveguide show great advances over traditional free space implementations in quantum information technology for significant advantages of low transmission loss, low cost, integrability and compatible with mature fiber communication systems. Many fundamental researches focus on generating of entangled sources and photonic quantum state engineering based on DSF [19-23] or SOI waveguide [24-26]. Though significant progresses have been achieved based on these two physical systems, no one reported on non-classical interference between two independent all fiber single photon sources based on DSF.

In this work, we report on non-classical interference between two independent single photons based on SFWM in two 300 m DSFs cooled by liquid nitrogen. We first give a general theoretical description of purity of photon pair generated in DSF, we deduce the optimized condition for high purity photon pair generation in DSF and numerically simulated the impact of spectral filtering on the photon pair purity. Then we perform the experiment with the guide of the theoretical prediction. We obtained a HOM interference visibility of 53.2±8.4% (82.9±5.3%) without (with) back ground coincidences subtracted. In addition, a higher coincidence to accidental coincidence ratio (CAR) of 131 is obtained by cooling the DSF with liquid nitrogen compared with our previous CAR at room temperature. Our results will be of great importance for scalable quantum information processing with all fiber based platform.

We'll first give a general theoretical description of photon pair generation based SFWM in DSF. The quantum state for the generated photon pair can be expressed as [11, 27]

$$\begin{aligned}|\Phi\rangle &= \iint d\omega_s d\omega_i f(\omega_s,\omega_i)\hat{a}^\dagger(\omega_s)\hat{a}^\dagger(\omega_i)|0\rangle \\ &= \sum_n \sqrt{g_n}\hat{A}_n^\dagger(\omega_s)\hat{B}_n^\dagger(\omega_i)|0\rangle\end{aligned} \quad (1)$$

Where $\hat{a}^\dagger(\omega_s)$ and $\hat{a}^\dagger(\omega_i)$ are creation operator of the signal and idler photon in frequency mode $\omega_s$ and $\omega_i$, respectively, $\hat{A}_n^\dagger(\omega_s)$, $\hat{B}_n^\dagger(\omega_s)$ are operators for Schmidt decomposition of modes $|\psi_n(\omega_s)\rangle, |\phi_n(\omega_i)\rangle$, where $\hat{A}_n^\dagger(\omega_s) = \int d\omega_s \psi_n(\omega_s)\hat{A}_n^\dagger(\omega_s)$, $\hat{B}_n^\dagger(\omega_i) = \int d\omega_i \phi_n(\omega_i)\hat{B}_n^\dagger(\omega_i)$, each of the mode has a weight given by the Schmidt eigenvalue $g_n$ [27]. The Schmidt decomposition makes it clear that heralding with one photon from the state $\hat{\rho}=|\Phi\rangle\langle\Phi|$ projects the remaining photon into state $\hat{\rho}_s = Tr_i(\hat{\rho})$, whose purity is defined by $p = Tr(\hat{\rho}_s^2) = \sum_n g_n^2$ and the purity is determined by the factorability of joint spectral amplitude (JSA) $f(\omega_s,\omega_i)$ [28], but in general, the Schmidt decomposition cannot be found analytically. Instead, one can calculate numerically and compute its singular value decomposition, the matrix analogue of the Schmidt decomposition. The purity of the reduced state is then given by the sum of the squares of the singular values, which are equal to the Schmidt magnitudes [29]. Hence a necessary condition for pure heralded photons is a factorable joint amplitude, $f(\omega_s,\omega_i) = f_s(\omega_s)f_i(\omega_i)$, with only one spectral mode present.

The JSA of the photon pair generated in DSF is determined by the pump envelope $\varepsilon(\omega_s + \omega_i)$

and the phase matching function $\Gamma(\omega_s,\omega_i)$ of the fiber, therefore $f(\omega_s,\omega_i) = \varepsilon(\omega_s+\omega_i)\Gamma(\omega_s,\omega_i)$. Assuming the pump laser has a Gaussian distribution with a bandwidth of $\sigma_p$, the pump envelope intensity can be expressed as $|\varepsilon(\omega_s+\omega_i)|^2 = \exp[-2(\frac{\omega_s+\omega_i-\omega_p}{\sigma_p})^2]$, the difference of a factor of 2 compared with SPDC in nonlinear crystal come from that two photons are annihilated in creating a signal and idler photon in SFWM [30]. The phase matching function for SFWM in DSF can be written as $|\Gamma(\omega_s,\omega_i)|^2 = [\sin c(\frac{\Delta k L}{2})]^2$, where $\Delta k = 2k_p - k_s - k_i + 2\gamma P$ is phase mismatch, here $k_{p,s,i}$ are the wave vector for the pump, signal and idler photon respectively, $\gamma$ is the third order nonlinear coefficient of the DSF and $P$ is the pump power. By approximating the phase matching function as Gaussian function $|\Gamma(\omega_s,\omega_i)|^2 \propto \exp(-2\alpha^2\Delta k^2 L^2)$, where $\alpha = 0.220$. By expanding the phase mismatching at the central frequencies $\omega_{p0}, \omega_{s0}, \omega_{i0}$ of the pump, signal and idler, the joint spectral intensity (JSI) can be approximated by [17]

$$S(\Delta\omega_s,\Delta\omega_i) = |f(\Delta\omega_s,\Delta\omega_i)|^2 \propto \exp[-2(\frac{1}{\sigma_p^2}+\alpha^2 L^2(k_p'-k_s')^2)\Delta\omega_s^2 - 2(\frac{1}{\sigma_p^2}+\alpha^2 L^2(k_p'-k_i')^2)\Delta\omega_i^2 - 4(\frac{1}{\sigma_p^2}+\alpha^2 L^2(k_p'-k_s')(k_p'-k_i'))\Delta\omega_s\Delta\omega_i] \quad (2)$$

Where $k_j' = dk/d\omega_j\big|_{\omega_j=\omega_{j0}}$ are the first deviations of the pump, signal and idler wave vectors, respectively. For a pure separable state, the condition $1+\sigma_p^2\alpha^2 L^2(k_p'-k_s')(k_p'-k_i') = 0$ should be fulfilled. For the DSF we used in our experiment, the optimized pump width should be 8 ps.

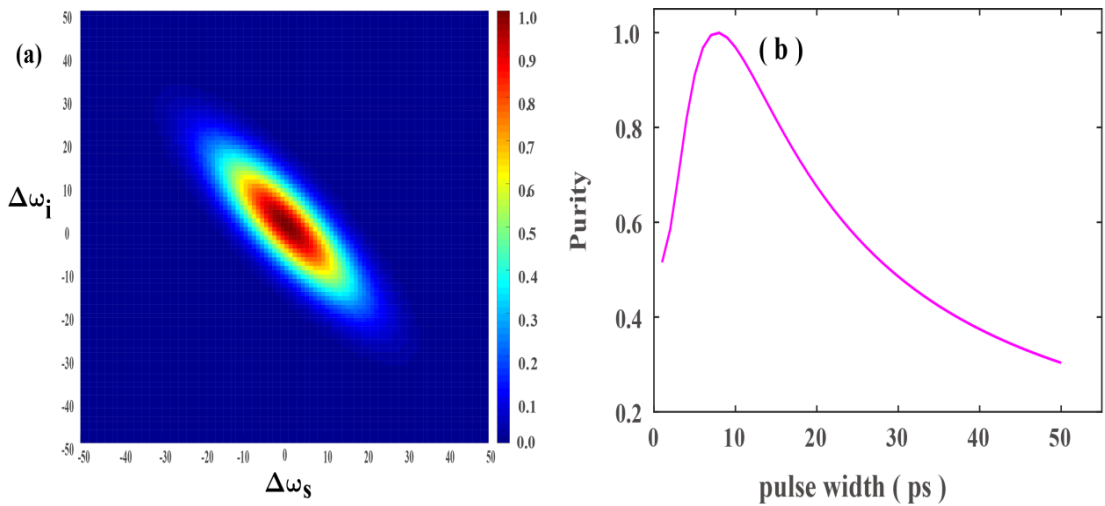

Figure 1. Numerical simulations of the JSI and purity. (a) JSI for signal and idler photon; (b) purity as a function of pump width.

By using our experimental parameters, we first numerically simulate the JSI and purity of the photon source and the influence of pump width on the purity. The results are showed in Fig. 1. Fig. 1(a) is JSI of the signal and idler photon based on our experimental parameters calculated using Eq. (2), the purity of our photon source is 0.5693. The influence of the pump pulse width on the purity is showed in Fig. 1(b), below the optimized pump width, the purity is increasing with the increasing of pump width, then further increasing of the pump pulse width, the purity is decreasing.

Then we experimental investigate the performance of our HOM interference between two independent all fiber single photon source. The experiment setup is shown in Fig.2. A homemade mode locked fiber laser source whose pulse width and repetition rate are 25 ps and 27.9MHz acts as pumps of photon sources. A tunable attenuator behind output of the pump laser is used to adjust the pump power of the photon sources, the broad band background fluorescence photons are filtered out by cascade 100GHz DWDM filters. The clean pump beam is divided into two equal parts by the 50:50 optical couplers to pump two 300m DSFs for generating independent photon sources, the two DSFs are cooled by liquid nitrogen. After two DSFs, the strong pump beam and noise photon from Raman scattering in orthogonal polarization are removed by 2 groups of 200GHz DWDM filters and fiber polarization rotator and polarizer. Then two 100GHz DWDMs are used for separating photons at correlated channel pairs. The relative delay between the two independent photons is adjusted by tunable fiber delay lines. Photons from the same channels are combined in a second beam splitter for interfering, the other two photons are used as trigger signals. The two triggered photons are detected by two free running InGaAs avalanched single photon detector ( APD1, APD4, ID Quanta, ID220, 20% detection efficiency, 3us dead time), outputs of detectors APD1 and APD4 are used to trigger two single photon detectors running in gated mode (APD2, APD3, Qsk, Hefei, China, 100MHz, free gating single photon detectors, 20% detection efficiency) for two photon coincidence measurement. Detection output signals from APD2 and APD3 are sent to our coincidence count device (Pico quanta, timeharp 260, 0.8 ns coincidence window) for four photon coincidence measurement.

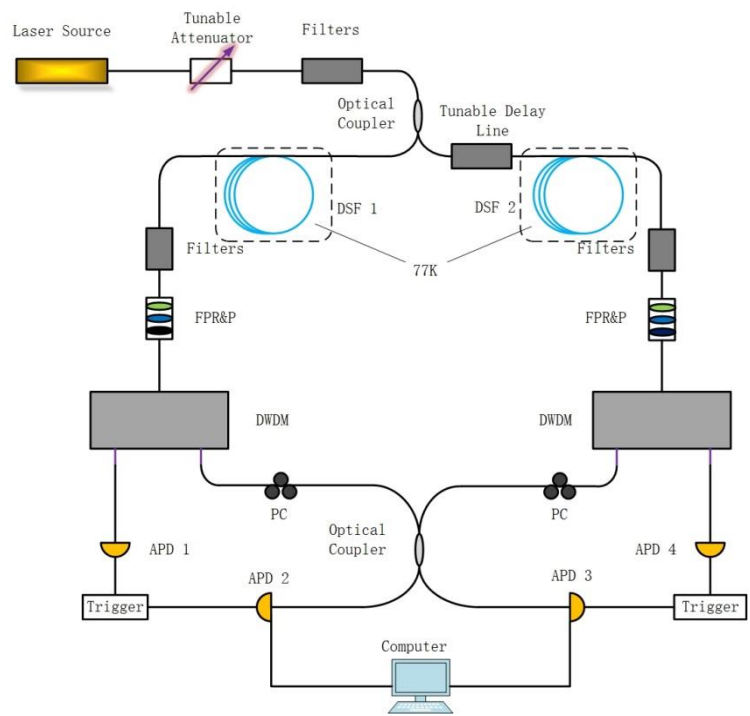

Figure 2. Experiment setup for HOM interference between two independent photon sources generated by SFWMs in two 300m DSFs. DSF: dispersion shifted fiber; APD1-4: avalanched photon detector. FPRP: fiber polarization rotator and polarizer; PC: polarization controller.

Before measuring HOM interference between independent single photon sources, we first characterize the property of single photon sources by measuring the CARs as a function of the pump power, the results are showed in Fig. 3(a). For relative low pump power, the CAR is dominated by the dark coincidences from the dark counts of the APDs, the CAR rises with increasing of pump power; which reaches maximum value of 131 at 23 µW. When the pump power is further increasing, multi-photon effect appears which limits further increasing of CARs, and the CARs decrease with the increasing of pump power. Then we perform four-fold HOM interference between signal1 (APD2) and signal2 (APD3), with idler1 (APD1) and idler2 (APD4) as heralded photons. The visibility without (with) dark coincidences subtracted is 57% ±8% (82.9 ±5.3%). The dark coincidences are measured by blocking one arm of the BS, and summing the two results together. The maximum visibility is determined by the following formula: [31]

$$V = Tr(\hat{\rho}_{s1}\hat{\rho}_{s2}) = \frac{Tr(\hat{\rho}_{s1}^2) + Tr(\hat{\rho}_{s2}^2) - \|\hat{\rho}_{s1} - \hat{\rho}_{s2}\|^2}{2} \quad (3)$$

Where $\hat{\rho}_{s1}, \hat{\rho}_{s2}$ are the density matrix of the two signal photons input on the BS, $\|\hat{\rho}_{s1} - \hat{\rho}_{s2}\|^2$ represents the indistinguishablity between the two photons. For identical photon pair source, the upper bound of the visibility is determined by the purity of the single photon pair source.

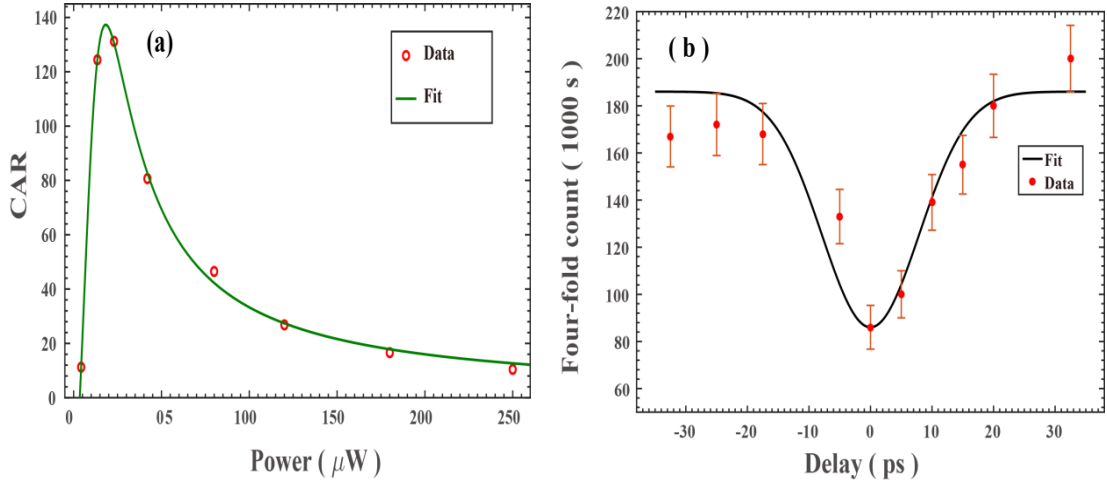

Figure. 3. The experimental results for CARs measurement for single photon source and four-fold HOM interference measurement between two photon sources. (a)CAR as a function of pump power; (b) four-fold coincidences in 1000 s as a function of the relative delay between the two single photon sources. The error bars are added by assuming Poisson distribution of the photon statistic.

To this point, we have given the theoretical and experimental investigations of HOM interference between two independent all fiber sources. There are a few remarks we want to point out: firstly the photon sources are not operated at the optimized pump pulse width for pure single photon generation, but narrow band 100GHz filters are used in the experiments to increase the purity of the sources for interference; Secondly, the free running detectors has relative low detection efficiency and much higher dark counts than nano-wire single photon detectors, if we have high performance nano-wire single photon detector, the experimental results will be greatly improved as the four-fold coincidences and dark coincidences scaling quadruplicate with the detection efficiency and dark count probability of a single detector. Thirdly, the realizing of HOM interference in this work is the key for future all fiber implementation of quantum teleportation, device independent quantum key distributions. Finally, though single 100GHz channel is used in our experiment, we can used other channels for multiplexing such interference processes to enhance the channel capacity in future quantum communications tasks.

In conclusion, we have realized HOM interference between two independent all fiber single photon sources, which solved the key problem for combining fiber based photon pair sources for realizing many important applications such as multi-photon entangled state engineering, quantum teleportation and device independent quantum key distribution. The free of free space alignment and integratiblity of fiber based platform will be very promising in advancing future scalable quantum information processing technology.

**Acknowledgement** We thank Dr. Hai-Ou Li for lending us liquid nitrogen for cooling the fiber, Dr. Wei Chen for lending us single photon detectors and Dr. Rui-Bo Jin for helpful discussions. This work was supported by the National Fundamental Research Program of China (Grant No. 2011CBA00200), the National Natural Science Foundation of China (Grant Nos. 11174271, 61275115, 61435011, 61525504) and the Fundamental Research Funds for the Central Universities.

**References**
[1] T. Nagata, R. Okamoto, J. L. O'Brien, K. Sasaki, and S.Takeuchi, Beating the standard quantum limit with four-entangled photons. Science 316, 726 (2007).
[2] J.-W. Pan, Z.-B. Chen, C.-Y. Lu, H. Weinfurter, A. Zeilinger, and M. Zukowski, Multiphoton entanglement and interferometry. Rev. Mod. Phys. 84, 777-838(2012).
[3] S. Pirandola, J. Eisert, C. Weedbrook, A. Furusawa, S. L. Braunstein, Advances in Quantum Teleportation. Nat. Photon. 9, 641-652 (2015).
[4] C. Wang, X.-T. Song, Z.-Q. Yin, S. Wang, W. Chen, C.-M. Zhang, G.-C. Guo, and Z.-F. Han, Phase-reference-free experiment of measurement-device-independent quantum key distribution. Phys. Rev. Lett. 115, 160502(2015).
[5] Y.-L. Tang, H.-L. Yin, S.-J. Chen, Y. Liu, W.-J. Zhang, X. Jiang, L. Zhang, J. Wang, L.-X. You, J.-Y. Guan, D.-X. Yang, Z. Wang, H. Liang, Z. Zhang, N. Zhou, X. Ma, T.-Y. Chen, Q. Zhang, and J.-W. Pan, Measurement-device-independent quantum key distribution over 200 km. Phys. Rev. Lett. 113, 190501(2014).
[6] C. Zhang, Y.-F. Huang, Z. Wang, B.-H. Liu, C.-F. Li, and G.-C. Guo, Experimental Greenberger-Horne-Zeilinger-type six-photon quantum nonlocality. Phys. Rev. Lett. 115, 260402(2015).
[7] X.-L. Wang, L.-K. Chen, W. Li, H.-L. Huang, C. Liu, C. Chen, Y.-H. Luo, Z.-E. Su, D. Wu, Z.-D. Li, H. Lu, Y. Hu, X. Jiang, C.-Z. Peng, L. Li, N.-L. Liu, Y.-A. Chen, C.-Y. Lu, J.-W. Pan. Experimental ten-photon entanglement. arXiv:1605.08547 [quant-ph].
[8] X.-C. Yao, T.-X. Wang, P. Xu, H. Lu, G.-S. Pan, X.-H. Bao, C.-Z. Peng, C.-Y. Lu, Y.-A. Chen and J.-W. Pan. Observation of eight-photon entanglement. Nat. Photon. 6, 225-228(2012).
[9] I. Afek, O. Ambar, Y. Silberberg, High-NOON states by mixing quantum and classical light.

Science 328, 879-8819(2010).

[10] R. Kaltenbaek, B. Blauensteiner, M. Zukowski, M. Aspelmeyer, and A. Zeilinger, Experimental interference of independent photons, Phys. Rev. Lett. 96, 240502 (2006).

[11] P. J. Mosley, J. S. Lundeen, B. J. Smith, P. Wasylczyk, A. B. U'Ren, C. Silberhorn, and I. A. Walmsley, Heralded generation of ultrafast single photons in pure quantum states. Phys. Rev. Lett. 100, 133601(2008).

[12] P. J. Mosley1, J. S. Lundeen, B. J. Smith and I. A. Walmsley, Conditional preparation of single photons using parametric downconversion: a recipe for purity. New J. of Phys. 10, 093011 (2008).

[13] M. Tanida, R. Okamoto, and S. Takeuchi, Highly indistinguishable heralded single-photon sources using parametric down conversion. Opt. Express 20, 15175-15285 (2012).

[14] J. Fulconis, O. Alibart, J. L. O'Brien, W. J. Wadsworth, and J. G. Rarity, Nonclassical interference and entanglement generation using a photonic crystal fiber pair photon source. Phys. Rev. Lett. 99, 120501 (2007).

[15] M. Halder, J. Fulconis, B. Cemlyn, A. Clark, C. Xiong, W. J. Wadsworth, and J. G. Rarity, Nonclassical 2-photon interference with separate intrinsically narrowband fibre sources. Opt. Express 17, 4670-4676 (2009)

[16] C. Soller, O. Cohen, B. J. Smith, I. A. Walmsley, and C. Silberhorn, High-performance single-photon generation with commercial-grade optical fiber. Phys. Rev. A 83, 031806(R) (2011).

[17] N. Bruno, A. Martin, T. Guerreiro, B. Sanguinetti, and R. T. Thew, Pulsed source of spectrally uncorrelated and indistinguishable photons at telecom wavelengths. Opt. Express 22, 17246-17253 (2014).

[18] R.-B. Jin, K. Wakui, R. Shimizu, H. Benichi, S. Miki, T. Yamashita, H. Terai, Z. Wang, M. Fujiwara, and M. Sasaki, Nonclassical interference between independent intrinsically pure single photons at telecommunication wavelength. Phys. Rev. A 87, 063801 (2013).

[19] X. Li, P. L. Voss, J. E. Sharping, and P. Kumar, Optical-Fiber Source of Polarization-Entangled Photons in the 1550 nm Telecom Band. Phys. Rev. Lett. 94, 053601(2015).

[20] H. Takesue and K. Inoue, Generation of polarization-entangled photon pairs and violation of Bell's inequality using spontaneous four-wave mixing in a fiber loop. Phys. Rev. A 70, 031802 (2004).

[21] H. Takesue and K. Inoue, Generation of 1.5- m band time-bin entanglement using spontaneous fiber four-wave mixing and planar light-wave circuit interferometers. Phys. Rev. A 72, 041804 (2004).

[22] S. X. Wang and G. S. Kanter, Robust multiwavelength all-fiber source of polarization-entangled photons with built-in analyzer alignment signal. IEEE J. of Selected Topics in Quantum electronics 15, 1733-1740(2009).

[23] Y. -H. Li, Z.-Y. Zhou, Z.-H. Xu, L.-X. Xu, B.-S. Shi, and G.-C. Guo, Multiplexed entangled photon sources for all fiber quantum networks. arXiv:1605.04701 [quant-ph].

[24] J. W. Silverstone, D. Bonneau, K. Ohira, N. Suzuki, H. Yoshida, N. Iizuka, M. Ezaki, C. M. Natarajan, M. G. Tanner, R. H. Hadfield, V. Zwiller, G. D. Marshall, J. G. Rarity, J. L. O'Brien and M. G. Thompson, On-chip quantum interference between silicon photon-pair sources. Nat. Photon. 8, 104-108(2014).

[25] C. Reimer, M. Kues, L. Caspani, B. Wetzel, P. Roztocki, M. Clerici, Y. Jestin, M. Ferrera, M. Peccianti, A. Pasquazi, B. E. Little, S. T. Chu, D. J. Moss, and R. Morandotti, Cross-polarized

photon-pair generation and bi-chromatically pumped optical parametric oscillation on a chip. Nat. Comunn. 6, 8236(2015).

[26] C. Reimer, M. Kues, P. Roztocki, B. Wetzel, F. Grazioso, B. E. Little, S. T. Chu, T. Johnston, Y. Bromberg, L. Caspani, D. J. Moss, R. Morandotti, Generation of multiphoton entangled quantum states by means of integrated frequency combs. Science 351, 1176-1180(2016).

[27] A. B. U'Ren, C. Silberhorn, K. Banaszek, I. A. Walmsley, R. Erdmann, W. P. Grice, and M. G. Raymer, Generation of pure-state single-photon wavepackets by conditional preparation based on spontaneous parametric downconversion. Laser Phys. 15, 146-161(2005).

[28]W. P. Grice, A. B. U'Ren, I. A. Walmsley, Eliminating frequency and space-time correlations in multiphoton states. Phys. Rev. A 64, 063815 (2001).

[29] C. K. Law, I. A. Walmsley, and J. H. Eberly, Continuous frequency entanglement: effective finite Hilbert space and entropy control. Phys. Rev. Lett. 84, 5304-5307 (2000).

[30] R.-B. Jin, R. Shimizu, K.Wakui, H. Benichi, and M. Sasaki, Widely tunable single photon source with high purity at telecom wavelength. Opt. Express 21, 10659-10666 (2013).

[31]J. Lee, M. S. Kim, G. Brukner, Operationally invariant measure of the distance between quantum states by complementary measurements. Phys. Rev. Lett. 91, 087902 (2003).